\documentclass[aps,prl,twocolumn,superscriptaddress,showpacs]{revtex4}

\usepackage{amsmath,amsfonts,amssymb,bbm,citesort,graphicx}

\newcommand{\eqn}[1]{\begin{equation} #1 \end{equation}} 
\newcommand{\aln}[1]{\begin{align} #1 \end{align}}       
\newcommand{\eq}[1]{(\ref{#1})}              
\newcommand{\wbar}{\overline}                
\renewcommand{\l}{\left}                     
\renewcommand{\r}{\right}                    

\begin{document}


\title{Statistical properties of one-dimensional random lasers}


\author{Oleg Zaitsev}
\email[E-mail: ]{oleg.zaitsev@uni-duisburg-essen.de}
\affiliation{Fachbereich Physik, Universit\"at Duisburg-Essen,
             Lotharstr.~1, 47048 Duisburg, Germany}

\author{Lev Deych}
\email[E-mail: ]{lev.deych@qc.cuny.edu}
\author{Vladimir Shuvayev}
\affiliation{Physics Department, Queens College of City University of
             New York, Flushing, NY 11367, U.S.A.}



\begin{abstract}

Statistical properties of a laser based on a one-dimensional disordered
superlattice open at one side are studied numerically. The passive normal modes
of the system are determined using the Feshbach projection technique. It is
found that the mode competition due to the spacial hole burning leads to a
saturation of the number of lasing modes with increasing pump rate. It is also
responsible for nonmonotonic dependence of intensities of lasing modes as
functions of pumping. Computed distributions of spectral spacing and intensity
statistics are in qualitative agreement with experimental results.

\end{abstract}


\maketitle


Random lasers~\cite{cao05} comprise a large variety of active radiative systems
that are based on disordered media. They possess one or more characteristics
that distinguish them from the traditional lasers such as absence of the
resonator, irregular electric field distributions and spectra, etc. A number of
problems can be formulated in connection with random lasers. One group of
questions is related to light localization in disordered systems with
amplification and absorption~\cite{cao05}. Another class of problems deals
with statistics of lasing modes~\cite{misi98,wu08,hack05,zait06,zait07}, e.g.,
distribution of their frequencies, intensities,~etc. From a methodological
point of view, extension of standard laser theory to random lasers is an
important open question, although some progress in this direction has been
made~\cite{misi98,hack05,deyc05a,ture06}.

Statistical properties of  random lasers were investigated in several
theoretical and experimental papers. Experimentally, distributions of the
spacial size of the modes and of the spacing between lasing frequencies in
porous GaP were reported in Ref.~\cite{mole07}. The spacing statistics were
also studied in lasers based on colloidal solutions~\cite{wu08}, where, in
addition, statistics of the emitted intensity were measured.

Theoretical studies of this issue were mainly performed for lasers with chaotic
resonators, which are similar, in certain aspects, to random lasers. In
Ref.~\cite{misi98} random-matrix theory was used to describe weakly open
chaotic cavities, where the average number of lasing modes for a given pump
strength and the probability distribution of the lasing thresholds were
derived. In Refs.~\cite{hack05,zait06,zait07} the average number of modes, its
fluctuations, and spectral spacing statistics were computed numerically,
particularly, for the case of cavities with overlapping resonances. Statistics
of light reflection by a random laser were studied in Ref.~\cite{been96}.

In this paper we study statistical properties of a particular model of random
lasers from first principles, without relying on random-matrix-type hypothesis
about statistical properties of cavity modes. Regular semiclassical multimode
laser theory~\cite{sarg74,hake84} was first applied to lasing in wave-chaotic
resonators in Ref.~\cite{misi98}. However, in the case of lasers based on
disordered structures, this theory has to be modified to take into account
several factors specific for these systems. First, in this case the openness of
the system is essential and should be properly incorporated into the theory. In
Ref.~\cite{vivi03} the modes of the open cavities were considered using
Feshbach projection technique~\cite{fesh62}, and this approach was applied
specifically to random laser systems in Ref.~\cite{hack05}. An alternative
approach based on so-called ``constant-flux modes'' was developed in
Ref.~\cite{ture06}. Second, the inhomogeneity of the background dielectric
constant introduces gain-induced coupling between modes, which is present
already in the linear approximation~\cite{deyc05a}, but can be enhanced due to
nonlinear hole burning effects~\cite{ture08}.

In this work we rederive Maxwell-Bloch lasing equations in the third order of
nonlinearity rigorously taking into account the openness of the system. These
equations are used for numerical statistical analysis of lasing from a
one-dimensional disordered structure. We apply the Feshbach method to compute
complex eigenfrequencies and wavefunctions of the structure under
consideration. It is our main conclusion that the correlation of the
wavefunctions critically affect lasing properties of the system. In particular,
they are responsible for saturation of the number of lasing modes and a
nonmonotonic behavior of lasing intensities with increasing pumping. Both these
effects are absent in the chaotic cavities described by random-matrix theory,
in which wavefunctions are not correlated~\cite{misi98,hack05}. A similar
nonmonotonic behavior of intensities was recently found in Ref.~\cite{ture08}
using a different approach. We also calculated statistical distributions of
spectral spacings and mode intensities, which show qualitative agreement with
experiment.

We consider a one-dimensional structure open at one side, which is
characterized by a real non-uniform refractive index $n(x)$ for $0 \leq x \leq
L$, $n=1$ for $x>L$, and an ideal mirror at $x=0$. Normal modes of this system
have a finite lifetime even in the absence of absorption. They can be found
using the system-and-bath approach~\cite{vivi03}. This approach requires
division of the entire system in a closed resonator and environment, which
interact through specified boundary conditions. In our formulation we designate
the region $0\leq x\leq L$ as a resonator with Dirichlet boundary condition at
$x=L-0$, then the bath region $x>L$ must be described by Neumann conditions at
$x=L+0$~\cite{vivi04}.

If  eigenfrequencies $\omega^c_\lambda$ and eigenfunctions $\psi^c_\lambda (x)$
of the closed resonator are known, then the true normal modes of the open
system are obtained by diagonalizing the non-Hermitian matrix
\eqn{ \Omega (\omega) = \Omega^c - i \pi W (\omega)\, W^\dag (\omega).
  \label{Omega}
 }
Here $\Omega^c$ is a diagonal matrix of the eigenfrequencies $\omega^c_\lambda$
and $W (\omega)$ is a column of coupling elements of the modes $\lambda$ with
the $\delta$-normalized bath modes $\sqrt{2/\pi} \cos[\omega (x - L)]$ (in the
system of units with the speed of light $c = 1$). Explicitly, $W_\lambda
(\omega) = (\psi^c_\lambda)' (L)/\sqrt{2 \pi \omega^c_\lambda\, \omega}$. The
eigenvalues of $\Omega (\omega)$, $\Omega_k (\omega) \equiv \omega_k (\omega) -
i \kappa_k (\omega)$, provide frequencies and decay rates of the normal modes.
Since $\Omega (\omega)$ is non-Hermitian we have to differentiate between left,
$|l_k (\omega) \rangle$, and right, $|r_k (\omega) \rangle$, eigenvectors,
which are biorthogonal. They can be used to define left and right
eigenfunctions $\psi_k^j (x; \omega) \equiv \sum_\lambda \psi^c_\lambda (x)
\langle \lambda | j_k (\omega) \rangle$, $j = l , r$, of the open system.
The electric field $E (x; \omega)$ in the frequency representation and other
relevant functions can be expanded in terms of these normal modes as $E (x;
\omega) = \sum_k E_k (\omega)\, \psi_k^r (x; \omega)$.

The gain medium is described by the polarization and the population difference
that interact with the classical field in the resonator. The atomic variables
can be eliminated perturbatively from the coupled equations of
motion~\cite{hake85}. Following Ref.~\cite{deyc05b} we carried out this
procedure in frequency domain, which is more convenient in our case than more
traditional time-domain consideration. The resulting equations for mode
amplitudes are
\begin{widetext}
\aln{
  &-i \l[\omega - \omega_k (\omega) + i \kappa_k (\omega) - i p D (\omega) \r]
  E_k (\omega) =
  \frac 1 2 p D (\omega) \int \frac {d \omega' d \omega''} {(2 \pi)^2}
  D^\parallel (\omega') \sum_{k_1, k_2, k_3} E_{k_1} (\omega - \omega') \notag
  \\
  &\times \l[ B_{kk_1k_2k_3} (\omega, \omega - \omega', \omega' - \omega'',
  -\omega'')\, D^* (-\omega'')\, E_{k_2} (\omega' - \omega'')\, E_{k_3}^*
  (-\omega'') \r. \notag \\
  &\l.+ B_{kk_1k_3k_2} (\omega, \omega - \omega', -\omega'', - \omega' -
  \omega'')\, D (-\omega'')\, E_{k_3} (-\omega'')\, E_{k_2}^* (- \omega' -
  \omega'') \r],
\label{field}
}
\end{widetext}
where the~r.h.s.\ contains nonlinear terms in the leading (third) order in the
field. In these equations, $p$ is a pump-rate parameter (uniform pumping is
assumed), $D (\omega) \equiv \l[ 1 - i (\omega - \nu)/\gamma_\perp \r]^{-1}$,
$D^\parallel (\omega) \equiv \l( 1 - i \omega / \gamma_\parallel \r)^{-1}$,
$\nu$ is the atomic-transition frequency, $\gamma_\perp$ and $\gamma_\parallel$
are polarization and population relaxation rates, and
\aln{
  &B_{k_1k_2k_3k_4} (\omega_1, \omega_2, \omega_3, \omega_4) \equiv
  L \int_0^L dx \, \l[\psi_{k_1}^l (x; \omega_1) \r]^*  \notag \\
  &\times \psi_{k_2}^r (x; \omega_2) \, \psi_{k_3}^r (x; \omega_3)
  \l[\psi_{k_4}^r (x; \omega_4) \r]^*
  \label{over}
}
denote the overlap integrals. The electric field (in the time representation)
is measured in units of $E_0 \equiv \sqrt{\hbar \gamma_\perp \gamma_\parallel /
8 \pi d^2 \nu}$, where $d$ is the atomic-transition dipole moment. The effect
of gain-induced linear mode coupling~\cite{deyc05a} was found to be small in
this system and is neglected in Eq.~\eq{field}. We do not take into account the
effects of linear gain and nonlinearity on the lasing frequencies approximating
them with solutions of equations $\omega_k (\omega) = \omega$, which will be
denoted as~$\omega_k$. In the slow-amplitude approximation, the mode amplitudes
$E_k (\omega)$ are assumed to be strongly peaked at $\omega_k$. The mode decay
rates, eigenvectors, and wavefunctions in this case can be taken at the
respective lasing frequencies, and the frequency arguments can be omitted. We
transform Eq.~\eq{field} to the time representation and obtain rate equations
for the intensities, ~$I_k \equiv |E_k (t)|^2$, in the form
\eqn{
  \dot I_k = 2 I_k \l(p |D_k|^2 - \kappa_k - p \sum_{k'} C_{kk'} I_{k'} \r), \;
  D_k \equiv D (\omega_k).
  \label{rate}
}
Here terms oscillating at beat frequencies have been ignored. The nonlinear
coupling between the modes $C_{kk'} = |D_{k'}|^2\, \text{Re}\, \l[B_{kkk'k'}
D_k \r]$ depends on the overlap integrals~\eq{over}. This is a simplified
expression for $C_{kk'}$, accurate when the population inversion is
time-independent. The correction due to population pulsations, $\Delta C_{kk'}
= (1/2)\, \text{Re}\, \l[B_{kk'kk'} D^\parallel (\omega_k - \omega_{k'})\,
D_k\, (D_k + D_{k'}^*)\r]$ ($k\ne k\prime$)~\cite{zait07}, can be disregarded
for typical situation $\gamma_\parallel \ll \gamma_\perp$. The r.h.s.\ of
Eq.~\eq{rate} has a transparent physical meaning of a balance between the gain,
damping, and nonlinear saturation that prevents the intensity from growing
indefinitely.

The amplitude of the emitted modes outside of the resonator,
$e_k^{\text{out}}$, is related to the internal mode amplitudes, $E_k$, by the
input-output relation~\cite{vivi03} with zero input: $e_k^{\text{out}} (t) = -i
E_k (t)\, W^\dag |r_k \rangle$. Using this relation in combination with natural
representation of matrix (\ref{Omega}) as a sum of Hermitian and anti-Hermitian
matrices, one can derive a physically transparent expression relating the
inside and outside intensities:
 \eqn{
  I_k^{\text{out}} = \kappa_k\, I_k / \pi.
  \label{Iout}
}

We model a one-dimensional disordered resonator by a superlattice $ABAB...$
consisting of $N_l$ layers of fixed width $a$. All layers $A$ have the same
refractive index~$n_A$ and for layers $B$ the index is drawn randomly from an
interval $(n_{\text{min}}, n_{\text{max}})$. In all our numerical examples the
parameters are $a = 1$, $N_l = L/a = 200$, $n_A = 1.5$, $n_{\text{min}} = 0.9$,
and $n_{\text{max}} = 1.3$ (the refractive index is defined relative to the
surrounding medium, where it is set to unity). This system is a
periodic-on-average structure with remnants of the bands of a periodic lattice
with $N_l/2$ elementary cells. While the band gaps are washed out by disorder
we can still define the boundaries of (quasi)bands as minima of the
disorder-averaged localization length (see inset of Fig.~\ref{fig2}). The
number of eigenmodes $\omega^c_\lambda$ for each band fluctuates from
realization to realization around $N_l/2$. We focus our simulations on the
frequency region of the third band $2.40 \leq \omega^c_\lambda \leq 3.61$, and,
respectively, restrict the basis of eigenvectors $\psi^c_\lambda (x)$ used to
construct matrix (\ref{Omega}) to the modes from this band. 

We computed numerically the stationary solutions ($\dot I_k = 0$) of
Eqs.~\eq{rate} which satisfy the condition $I_k (p) \geq 0$ while continuously
increasing pumping from zero.  It should be understood that this method does
not allow us to detect situations when a mode loses its stability without
passing through zero intensity.

\begin{figure}
  \vspace*{.1cm}
  \centering{\includegraphics[width= .76\linewidth, angle=0]{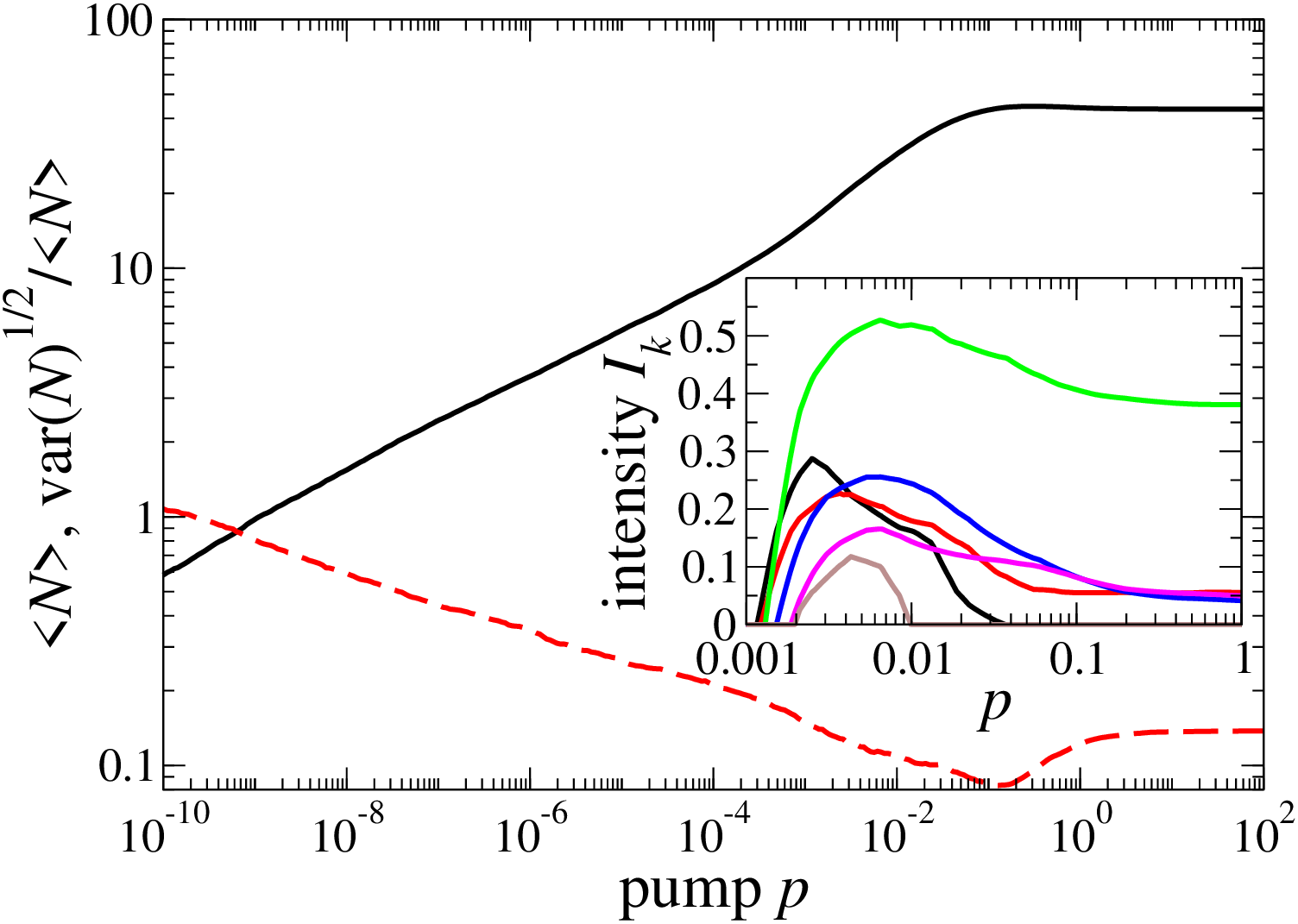}}
  \caption{(Color online) Average number of lasing modes $\langle N \rangle$
  (solid line) and its fluctuations $\sqrt{\text{var} (N)}/\langle N \rangle$
  (dashed line), as a function of the pump rate~$p$. For small $p$, $\langle
  N\rangle \propto p^{0.19}$ and $\sqrt{\text{var} (N)}/\langle N \rangle
  \propto p^{-0.12}$. Inset: intensities~$I_k (p)$ of several lasing modes
  excited in the intermediate pump range, for one disorder realization.
  Parameters: $\nu = 3.005$ and $\gamma_\perp = 0.242$.}
  \vspace{-.2cm}
  \label{fig1}
\end{figure}

\begin{figure}
  \vspace*{.1cm}
  \centering{\includegraphics[width= .76\linewidth, angle=0]{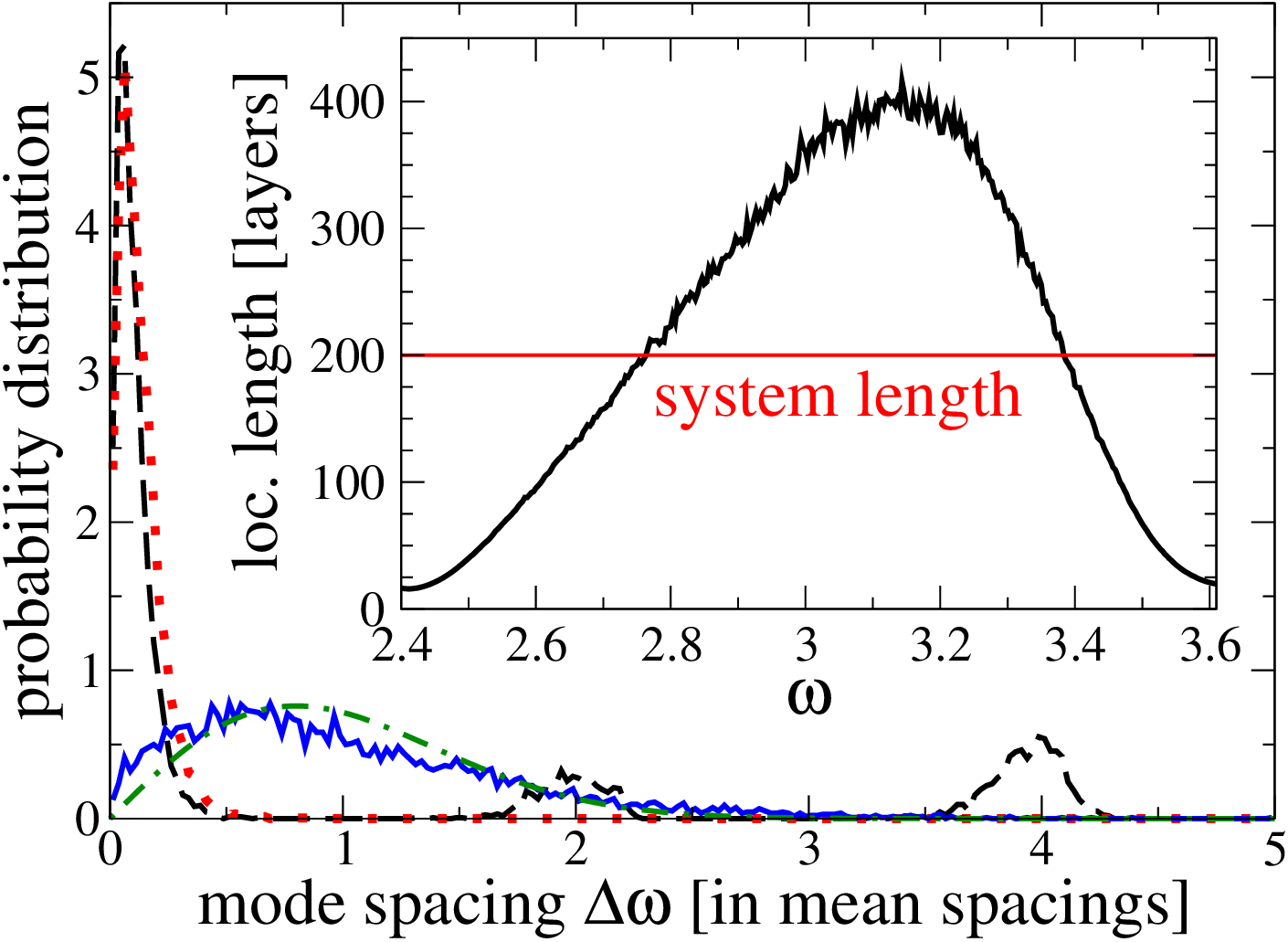}}
  \caption{(Color online) Probability distribution of the spectral spacing
  between lasing modes ($\gamma_\perp = 0.0121$, $p = 10^{-3}$). Parameters and
  properties for the dashed, dotted, and solid lines, respectively: atomic
  frequency $\nu = 3.005$, $2.405$, $2.405$; modes collected from the frequency
  interval $(2.400, 3.610)$, $(2.400, 3.610)$, $(2.400, 2.521)$; determined
  mean spacing $\wbar{\Delta \omega} =  0.27$, $0.18$, $0.019$. Dash-dotted
  line: Wigner surmise (see the text). Inset: average localization length (in
  units of the layer width) in the third quasiband, as a function of
  eigenfrequency~$\omega$. Horizontal line indicates current system size.}
  \vspace{-.3cm}
  \label{fig2}
\end{figure}

The results presented below are obtained with 1700 realizations of disorder.
Figure~\ref{fig1} shows the pump dependence of the average number of modes
$\langle N \rangle$ and its relative fluctuations $\sqrt{\text{var}
(N)}/\langle N \rangle$. The atomic frequency $\nu$ is chosen at the middle of
the band and the gain width $\gamma_\perp$ is equal to 20\% of the band width.
An important effect revealed in this figure is the saturation of $\langle N
\rangle$ for large~$p$ at the level of 44~modes. This phenomenon is not an
artifact of the finite basis, which contains 100 available modes. The
saturation appears as a result of the nonmonotonic pump dependence of
intensities~$I_k (p)$ and mode suppression (see the inset). The nonmonotonic
pump dependence was observed numerically in two-dimensional disordered lasers
as well~\cite{ture08}. Since in systems with uncorrelated wavefunctions, such
as two-dimensional chaotic resonators, where $B_{kkk'k'} = 1 + 2 \delta_{kk'}$,
mode suppression does not occur~\cite{misi98}, we relate the origin of the
saturation effect to statistical correlations between modes manifested via the
overlap integrals~$B_{kkk'k'}$. This effect should not be confused with
saturation of the number of lasing modes in the regime of strong
localization~\cite{jian00}, where saturation is due to ``exhaustion'' of the
number of available nonoverlapping localized modes. In the situation considered
here the system is in a truly multimode regime with overlapping modes and the
saturation is a result of a nontrivial interplay between effects of self- and
cross-saturation.

Before saturation the variance of the number of modes behaves as $\text{var}(N)
\propto \langle N \rangle^{0.74}$. Deviation from this behavior marks
transition to the regime of strong nonlinear mode competition.

Figure~\ref{fig2} displays probability distributions of spacing~$\Delta \omega$
between frequencies of the neighboring lasing modes. If the atomic frequency
$\nu$ is at the band center and $\gamma_\perp$ is rather small (1\% of the band
width, in this example), then the distribution has three maxima (dashed line).
This behavior results from the existence of well localized modes at the both
band edges that despite experiencing a small gain still can lase due to their
long lifetimes. As a result we have three well separated groups of lasing modes
giving rise to three maxima in the spacing distribution. If the gain is
centered at the band edge (dotted line), most of the lasing modes come from
that edge, but modes with extremely small~$\kappa_k$ at the opposite edge can
be excited as well. This produces a second maximum at the full band width
(outside of the plot range).

The solid line in Fig.~\ref{fig2} presents local spacing distribution for modes
taken from a narrow spectral strip (10\% of the band width) at the band edge,
$\nu$~being within the strip. This distribution displays mode repulsion
(vanishes in the limit $\Delta \omega \to 0$) and is similar to the Wigner
surmise $P_W (\Delta \omega) = (\pi \Delta \omega/ 2) \exp (- \pi \Delta
\omega^2/ 4)$ (dash-dotted line), which approximates the spacing distribution
in passive closed chaotic systems. The repulsion occurs because the length $L$
is short enough, so that even states with localization length of $L/10$ can
spatially overlap. Moreover, two nonoverlapping modes may have close
frequencies, but one of them will be localized closer to the opening and
suppressed for moderate pump strength.

\begin{figure}
  \vspace*{.1cm}
  \centering{\includegraphics[width= .76\linewidth, angle=0]{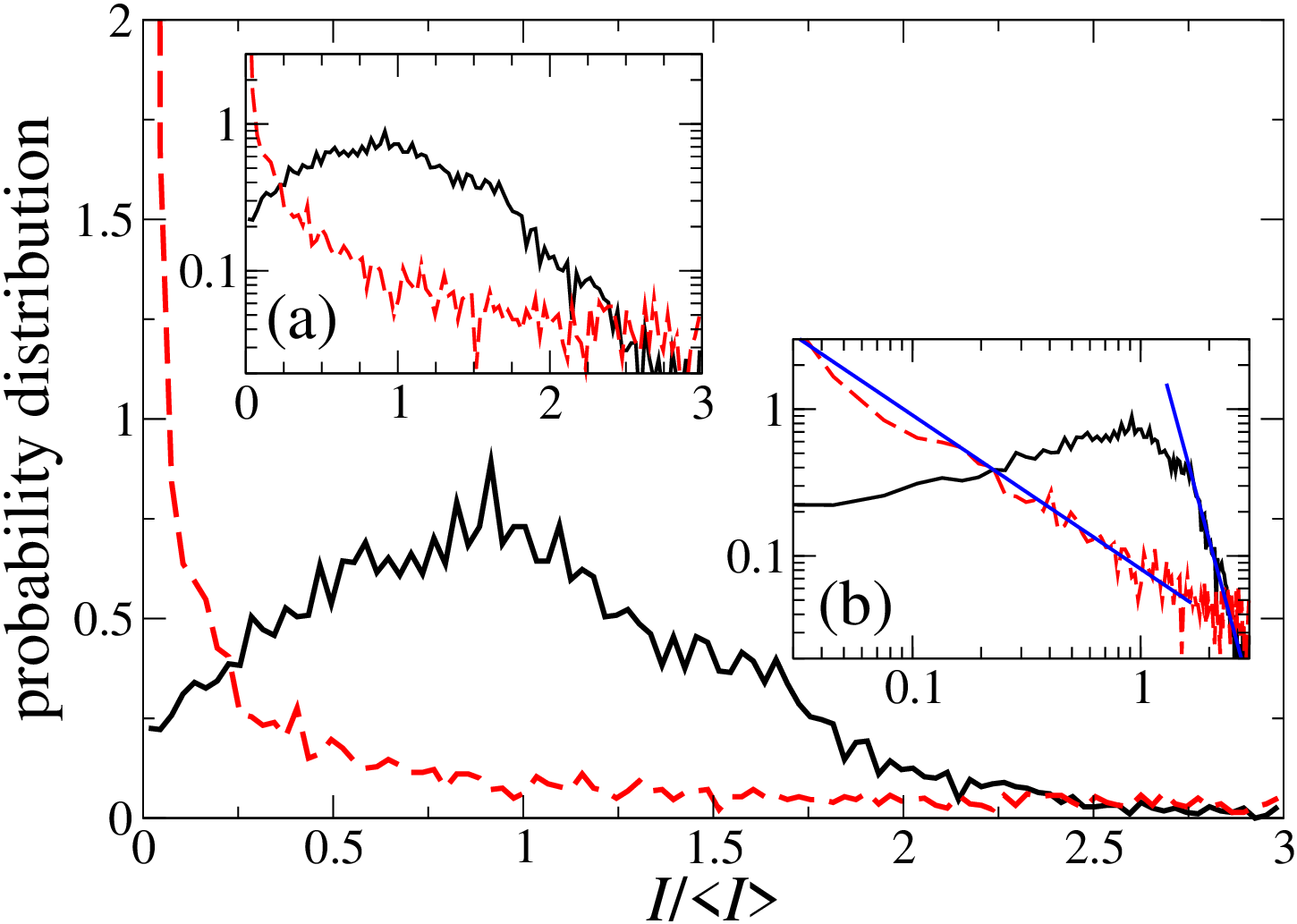}}
  \caption{(Color online) Probability distributions $P (I/\langle I\rangle)$ of
  internal mode intensities~$I_k$ (solid line) and output
  intensities~$I_k^{\text{out}}$ (dashed line). Parameters: $\nu = 2.405$,
  $\gamma_\perp = 0.242$, $p = 10^{-3}$; modes are collected from the frequency
  interval $(2.400, 2.521)$. Mean values: $\langle I_k \rangle = 0.135$,
  $\langle I_k^{\text{out}} \rangle = 1.78 \times 10^{-6}$. Insets:
  (a)~semilogarithmic plots; (b)~logarithmic plots with the fits $P(x) = 6.6\,
  x^{-5.7}$ (internal) and $P(x) = 0.082\, x^{-1.0}$ (output), where $x \equiv
  I/\langle I\rangle$.}
  \vspace{-.2cm}
  \label{fig3}
\end{figure}

Probability distributions for internal and output intensities of lasing modes
are shown in Fig.~\ref{fig3}. The output distribution has a singularity at zero
intensity, whose existence is related to a very broad (several orders of
magnitude) distribution of the decay rates of the modes in our system. Indeed,
the modes with smaller decay rates~$\kappa_k$ are preferentially excited, but
have lower output intensities according to Eq.~\eq{Iout}. Both distributions
have approximately a power-law asymptotic behavior at large intensities, with
different exponents (see the insets).

It is interesting to compare our numerical results for the spectral spacing
(Fig.~\ref{fig2}) and the intensity distributions (Fig.~\ref{fig3}) with
available experimental data. The spacing distribution in porous GaP was found
to be well approximated by the Wigner surmise~\cite{mole07}, while in colloidal
solutions of $\text{TiO}_2$ particles the mode repulsion was not of Wigner
type~\cite{wu08}. At the same time, the power-law distribution of intensities
of lasing modes found in Ref.~\cite{wu08} agrees  with our simulations.

In conclusion, we studied numerically a model of disordered laser based on a
one-dimensional open resonator. The passive normal modes of the system were
determined self-consistently using the Feshbach projection technique. The
intensities of lasing modes were found from the rate equations within the
semiclassical third-order laser theory. Mode competition, as a consequence of
the spacial hole burning, leads to nonmonotonic pump dependence of intensities
and mode suppression. The number of lasing modes saturates with increasing pump
rate. The local spectral spacing distribution shows a Wigner-like mode
repulsion. Globally, the distribution can have several maxima due to the
quasiband structure of the spectrum. Distributions of the mode intensities have
a power-law asymptotic tail. Output intensities are distributed over several
orders of magnitude, reflecting the spread of radiative lifetimes of normal
modes.

\begin{acknowledgments}
We would like to thank Fritz Haake for helpful discussions. Financial support
by the Deutsche Forschungs\-gemein\-schaft via the SFB/TR12 (O.Z.) and by
AFOSR via grant F49620-02-1-0305, as well as support by PCS-CUNY grants (L.D.)
is acknowledged.
\end{acknowledgments}



\end{document}